\documentclass[12pt]{article}
\usepackage{graphicx}
\usepackage{tikz}
\usepackage{algpseudocode}
\usepackage{algorithm}
\usepackage{natbib}
\usepackage{xcolor}
\usepackage{mathtools}
\usepackage{setspace}
\usepackage{latexsym}
\usepackage{tabulary}
\usepackage{booktabs,array,multirow}
\usepackage{amsfonts,amsmath,amssymb}
\usepackage{url}
\usepackage{hyperref}
\usepackage{authblk}
\usepackage{geometry}

\hypersetup{colorlinks=false,pdfborder={0 0 0}}

\AtBeginDocument{\DeclareGraphicsExtensions{.pdf,.PDF,.eps,.EPS,.png,.PNG,.tif,.TIF,.jpg,.JPG,.jpeg,.JPEG}}

\usepackage[utf8]{inputenc}
\usepackage[english]{babel}

\newcommand{\pro}{^\prime}
\newcommand{\obs}{_{\text{\tiny obs}}}

\newcommand{\logcs}{\alpha}
\newcommand{\tcell}{T_\text{c}}
\newcommand{\unif}[1]{\text{Uni}\left( #1 \right)}
\newcommand{\transp}{^T}

\title{ABC of the Future}
\date{}

\author[1]{Henri Pesonen}

\author[3]{Umberto Simola}

\author[1]{Alvaro K\"ohn-Luque}

\author[4]{Henri Vuollekoski}

\author[1]{Xiaoran Lai}

\author[1, 2]{Arnoldo Frigessi}

\author[4, 5]{Samuel Kaski}

\author[6]{David T. Frazier}

\author[7]{Worapree Maneesoonthorn}

\author[6]{Gael M. Martin}

\author[1, 3, 8]{Jukka Corander}

\affil[1]{Oslo Centre for Biostatistics and Epidemiology, University of Oslo, Oslo, Norway}

\affil[2]{Oslo Centre for Biostatistics and Epidemiology, Oslo University Hospital, Oslo, Norway}

\affil[3]{Helsinki Institute of Information Technology, Department of Mathematics and Statistics, University of Helsinki, Finland}

\affil[4]{Helsinki Institute of Information Technology, Department of Computer Science, Aalto University, Finland}

\affil[5]{Department of Computer Science, University of Manchester, UK}

\affil[6]{Department of Econometrics \& Business Statistics, Monash University, Australia}

\affil[7]{Melbourne Business School, the University of Melbourne, Australia}

\affil[8]{Parasites and Microbes, Wellcome Sanger Institute, UK}

\begin{document}

\maketitle
\selectlanguage{english}

\begin{abstract}Approximate Bayesian computation (ABC) has advanced in two decades from a seminal idea to a practically applicable inference tool for simulator-based statistical models, which are becoming increasingly popular in many research domains. The computational feasibility of ABC for practical applications has been recently boosted by adopting techniques from machine learning to build surrogate models for the approximate likelihood or posterior and by the introduction of a general-purpose software platform with several advanced features, including automated parallelization. Here we demonstrate the strengths of the advances in ABC by going beyond the typical benchmark examples and considering real applications in astronomy, infectious disease epidemiology, personalised cancer therapy and financial prediction. We anticipate that the emerging success of ABC in producing actual added value and quantitative insights in the real world will continue to inspire a plethora of further applications across different fields of science, social science and technology.

\textbf{keywords} --- Approximate Bayesian computation, Likelihood-free inference, Simulator-based inference, Bayesian inference
\end{abstract}

\section{Introduction}

From its humble beginnings over two decades ago, approximate Bayesian computation, or ABC for short, has more recently been met with ever-increasing excitement and is now regarded as one of the most transformative ideas in the statistical sciences. 
For in-depth reviews of the history of ABC, its theoretical underpinnings and the most recent developments, see \citep{marin2012, sisson2018, beaumont2019}. 
In particular, deeper theoretical insights into the large-sample behavior of ABC inference have removed
some of the main doubts regarding its statistical validity, and the field is moving rapidly towards a unified understanding of the key aspects that impinge on the asymptotic behavior of ABC approximations.  \citep{fearnhead2012, marin2014,  green2015, lintusaari2017, frazier2018, LF2016a, LF2016b, frazier2020model}.
Nevertheless, the generic application potential of ABC and other likelihood-free inference (LFI) methods has been held back by the computational requirements of its standard inference algorithms and the lack of a suitable all-purpose software implementation. With the advent of more efficient inference strategies adopted from the field of machine learning \citep{gutmann2016bayesian, gutmann2018, lueckmann2018, thomas2019, kokko2019, cranmer2019, grazian2019, papamakarios2019} and  software platforms such as Engine for likelihood-free inference (ELFI) \citep{Lintusaari2018}, ABCpy \citep{Dutta:2017:AUE:3093172.3093233}, BSL \citep{an2019bsl} and sbi \citep{tejero-cantero2020sbi}, to name a few, the immediate prospect of both using and updating the ABC/LFI toolkits for challenging real-world applications certainly looks brighter.

For example, Bayesian optimization for likelihood-free inference (BOLFI), has been shown in several benchmark examples to speed up ABC inference by $3$--$4$ orders of magnitude \citep{gutmann2016bayesian}, and multiple successful applications of it beyond typical benchmarks used in the statistical literature have emerged. These include applications in very diverse research fields, such as inverse reinforcement learning for cognitive user interface models \citep{kangasraasio2017}, brain task interleaving modeling \citep{gebhart2020} and more general computational models of cognition \citep{kangasraasio2019}, perturbation modeling and selection in bacterial populations \citep{corander2017}, direct dark matter detection \citep{simola2019machine}, pathogen outbreak modeling  \citep{lintusaari2019}, sound source localisation \citep{forbes2021}, passenger flow estimation in airports \citep{ebert2021}, and the modeling of the genetic components that control the transmissibility of pathogens \citep{shen2019}. 
To inspire further methodological development, software engineering and dissemination of ABC and other LFI methods, we present here an array of real applications and discuss both the benefits and challenges that lie ahead for this exciting subfield of statistics.

\section{Statistical Methodology} \label{sec.stat.met}
\subsection{Preliminaries}

We briefly introduce the main concepts and the notation used in the following sections, using the simplest form of ABC algorithm for the purpose of illustration. More detailed description of the methods used in the examples, i.e.~approximate Bayesian computation--population Monte Carlo (ABC-PMC) \citep{BeaumontEtAl2009} and BOLFI \citep{gutmann2016bayesian} can be found in the following sections.

{\color{black}Bayesian inference is based on calculating the posterior distribution 
\begin{equation}
    p(\theta \mid y\obs) \propto p(y\obs \mid \theta)p(\theta)
\end{equation} 
of the parameters $\theta$ given the observed data $y\obs$. The commonly employed methods of} {\color{black}conducting inference} {\color{black} based on the posterior (e.g. optimization, importance sampling, Markov chain Monte Carlo (MCMC))} {\color{black} all require pointwise evaluation of the likelihood, $p(y\obs \mid \theta)$, at any $\theta$.} {\color{black} ABC provides an inferential framework for situations where the likelihood function} {\color{black} is not available, or whose evaluation is too computationally challenging,  by replacing likelihood \textit{evaluation} with \textit{simulation} of the data--generating process, where the latter task is often still feasible even when the former is not.} 

The rejection ABC algorithm {\color{black}in Algorithm \ref{alg1:rejABC.alg}} \citep{TavareEtAl1997,PitchardEtAl1999} is the basic formulation of an ABC method. Assuming that the simulator parameter $\theta \in \mathbb R^p$ is the target of the statistical inference, the rejection ABC algorithm produces $N$ independent samples from an approximate ABC posterior distribution
\begin{equation}
\pi_{\epsilon}(\theta \mid y\obs) \propto \int  {\color{black} p}(y\pro \mid \theta){\color{black} p}(\theta)\mathbb I_{A_{\epsilon,y\obs}}(y\pro)\text{d}y\pro, \label{ABCposterior}
\end{equation}
where {\color{black}$y\pro$ is artificial data simulated from the generative model}, $\mathbb I_{A_{\epsilon,y\obs}}(\cdot)$ is the indicator function for the set $A_{\epsilon,y\obs} = \{y\pro \mid d(y\obs, y\pro) \leq \epsilon\}$, {\color{black} which is defined via a distance metric $d(\cdot, \cdot)$ and a threshold parameter $\epsilon$.}

The ABC posterior as defined by \eqref{ABCposterior} is not conditioning on the data exactly, but on a set of artificial data following the distribution of the generative model that is within a tolerance $\epsilon$ from the observed data, as determined by the difference $d(y\obs, y\pro)$ between observed and artificial data.
Because the relative volume of  $A_{\epsilon,y\obs}$ becomes vanishingly small when the dimension of the data $y\obs$ increases, sampling-based algorithms such as rejection ABC reduce the sample space in order to perform adequately. Traditionally, this is done by defining a set of summary statistics $s(\cdot)$. Ideally, the summarising function would be a sufficient statistic, but this is rarely available in problems with intractable likelihoods. In practice, $s(\cdot)$ are chosen according to a number of different principles, aimed to maximize the informativeness of the summaries in some sense  \citep{JoyceMarjoram2008, Blum2010, DPF2011, fearnhead2012, DPL2015, martin2019auxiliary}. 
The core of {\color{black}Algorithm \ref{alg1:rejABC.alg}} is simple and straightforward to implement in most programming languages. 

\begin{algorithm}[htbp]
\caption{Rejection ABC algorithm for $\theta$}
\begin{algorithmic}
 \For{$i = 1, \ldots, N$}
    \State $d\pro = \infty$
    \While{$d\pro > \epsilon$}
	\State Sample from the prior, $\theta\pro \sim {\color{black} p}(\theta)$
 	\State Simulate from the generative model, $y\pro \sim {\color{black} p}\left(y\mid \theta\pro\right)$
 	\State Calculate distance, $d\pro = d(s(y\obs), s(y\pro))$
 	\EndWhile
 	\State Set $\theta^{(i)} = \theta\pro$
 \EndFor
\end{algorithmic} \label{alg1:rejABC.alg}
\end{algorithm}

\subsection{ABC--PMC algorithm} \label{intro.abcpmc}

Although the rejection ABC algorithm is still being used frequently as a comparison method in the ABC and LFI literature, there are few applications where it would not be beneficial to instead take advantage of more sophisticated versions of this basic algorithm \citep{BeaumontEtAl2009, Blum2010, CsilleryEtAl2010, DrovandiEtAl2011, marin2012, DelMoralEtAl2012, clarte2019component, rodrigues2019likelihood, simola2020adaptive}. 
The ABC-PMC approach by  \cite{BeaumontEtAl2009} is an extension of the rejection ABC  algorithm based on importance sampling, and aims to improve the efficiency of the procedure by retrieving a sequence of intermediate  distributions. The steps of the method are summarized in Algorithm~\ref{alg2:ABC--PMC.alg}. The first iteration of the ABC--PMC algorithm corresponds to the four steps of the basic rejection ABC algorithm with $\epsilon_1$.  Starting from the second iteration $t \geq 2$, 
a particle is sampled with replacement from the set of importance weighted particles at iteration $t-1$, it is moved using e.g.~a Gaussian kernel, and accepted or rejected based on $\epsilon_t < \epsilon_{t-1}$.   The importance weight for particle $J = 1, \ldots, N$ at iteration $t$ is
\begin{equation*}
W_t^{(J)} \propto {\color{black} p}(\theta_{t}^{(J)})/\sum_{K = 1}^N W_{t-1}^{(K)} \phi\left[\tau_{t-1}^{-1/2}\left(\theta_{t}^{(J)} - \theta_{t-1}^{(K)}\right)\right],
\end{equation*}
where $\phi(\cdot)$ is a multivariate Gaussian kernel with mean 0 and identity covariance, and $\tau_{t-1}$ is  twice the (weighted) sample covariance of the particles from iteration $t-1$, as recommended in \cite{BeaumontEtAl2009}.

\begin{algorithm}[!ht]
\caption{ABC--PMC algorithm for $\theta$}
\begin{algorithmic}
\State Set $\epsilon_1 > \ldots > \epsilon_T$
\If{$t = 1$}
\For{$J = 1, \ldots, N$}
 \While{$d_1^{(J)} > \epsilon_1$}
	\State Propose $\theta^{(J)} \sim {\color{black} p}(\theta)$
 	\State Generate  $y\pro \sim {\color{black} p}\left(y\mid \theta^{(J)}\right)$
 	\State Calculate $d_1^{(J)} = {\color{black} d}(s(y\obs), s(y\pro))$
 \EndWhile
 	\State Set weight $W_1^{(J)} = N^{-1}$
 \EndFor
 \ElsIf{$2 \leq t \leq T$}
  \State Set $\tau_{t} = 2 \cdot \text{cov} \left( \{\theta_{t-1}^{(J)},W_{t-1}^{(J)}\}_{J=1}^{N}\right)$
 \For{$J = 1, \ldots, N$}
  \While{$d_t^{(J)}> \epsilon_t$}
	\State Select $\theta_t^*$ from $\theta_{t-1}^{(J)}$ with probabilities $\left\{W_{t-1}^{(J)}/\sum_{K = 1}^NW_{t-1}^{(K)}\right\}_{J = 1}^N$
	\State Propose $\theta_t^{(J)} \sim N(\theta_t^*, \tau_{t})$
	\State Generate  $y\pro \sim {\color{black} p}\left(y \mid \theta_{t}^{(J)}\right)$
 	\State Calculate $d_t^{(J)} = {\color{black} d}(s(y\obs), s(y\pro))$
 \EndWhile
 	\State Set weight $W_t^{(J)} \propto {\color{black} p}(\theta_t^{(J)})/\sum_{K = 1}^N W_{t-1}^{(K)} \phi\left[\tau_{t-1}^{-1/2}\left(\theta_{t}^{(J)} - \theta_{t-1}^{(K)}\right)\right]$
 \EndFor
 \EndIf
\end{algorithmic} \label{alg2:ABC--PMC.alg}
\end{algorithm}

As a final remark on the ABC--PMC algorithm, both the total number of iterations $T$ and the series of decreasing tolerances $\epsilon_1 > \epsilon_2 > \cdots > \epsilon_T$ must be selected in advance by the researcher, which might have an impact on the computational performance of the procedure, as well as on the attainability of a suitably accurate approximation of the exact posterior distribution \citep{Silk2013, simola2020adaptive}. For this reason, rather than defining in advance the series of decreasing tolerances, other choices are possible. In particular, adaptively selecting the tolerance for some iteration $t$  based on a previously defined quantile of the distances of the accepted particles from the previous iteration $t-1$ improves the efficiency of the algorithm in terms of how many times the forward model is used \citep{Lenormand2013, WeyantEtAl2013, IshidaEtAl2015, Cisewski-Kehe:2019aa, simola2020adaptive}. For these reasons also a tuning step is often necessary in order to balance the trade--off between computational efficiency and the reliability of the inferential results.

\subsection{BOLFI algorithm} \label{intro.bolfi}

Although potentially orders of magnitude more efficient than the basic rejection ABC algorithm, ABC--PMC is based on the idea of identifying relevant regions of the parameter space  by finding proposal distributions such that the corresponding simulated datasets are similar within a tolerance to the observed dataset, according to pre-defined summary statistics. As often only weak information is available in advance about what regions are relevant, this requires a large number of datasets to be simulated through the forward model.  \cite{Lintusaari2018} notes, for example, that the number of simulated datasets required in the implementation of a sequential ABC sampler like Algorithm~\ref{alg2:ABC--PMC.alg} is usually at least of order $10^6$. If model simulation itself is heavy, the total computational cost of the sequential algorithm becomes very high.

To reduce the need for a large number of potentially costly model simulations, active learning methods adapt the querying process according to different strategies. BOLFI uses Bayesian optimization \citep{gutmann2016bayesian} to construct iteratively a probabilistic surrogate model 
for the relationship between parameters $\theta\pro$ and discrepancies $d(\theta\pro) = {\color{black} d}(s(y\pro), s(y\obs))$ using the growing evidence set $\mathcal{E}_{t}$, which consists of pairs $\{(\theta_{i}, d(\theta_{i}))\}_{i=1}^t$.
The original formulation of BOLFI uses a Gaussian process (GP) as the surrogate model for the discrepancy function, and new evidence  $\{(\theta_{t+1}, d(\theta_{t+1}))\}$ is sampled from relevant regions of the parameter space. 
Relevant regions are determined to be parts of the space where the discrepancy is small. 
The probabilistic model is defined as $d(\theta) \mid \mathcal{E}_t  \sim GP(\mu_t(\theta), v_t(\theta) + \sigma^2)$, where $GP$ is a Gaussian process with mean and variance functions, 
\begin{align}
\mu_t(\theta) &= k_t(\theta)\transp K_t^{-1}[d(\theta_1), \ldots, d(\theta_t)]^T \\
v_t(\theta) + \sigma^2 &= k(\theta, \theta) - k_t(\theta)\transp K_t^{-1}k_t(\theta) + \sigma^2.
\end{align}
The vector $k_t(\theta) = \begin{bmatrix} k(\theta, \theta_1), \ldots, k(\theta, \theta_t)  \end{bmatrix}\transp$ and matrix $K_t = [k(\theta_i, \theta_j)] \in \mathbb{R}^{t \times t}$ are defined via covariance functions  $k(\theta\pro, \theta^{\prime\prime})$.

A common choice for a covariance function is  the squared exponential covariance function
\begin{align}
    k(\theta\pro, \theta^{\prime\prime}) &= \sigma_f^2\exp\left( \sum_{j=1}^{d} \frac{1}{\lambda_j^2}(\theta\pro_j - \theta^{\prime\prime}_j)^2 \right),
\end{align}
where parameters $\theta\pro_j$ and $\theta^{\prime\prime}_{j}$ are the $j$th elements of $\theta\pro$ and $\theta^{\prime\prime}$.
When fitting a GP to the evidence set, the hyperparameters $\sigma$, $\sigma_f$,  and $\lambda_j$ can be optimized iteratively \citep{rasmussen2006}. In practice, it is not necessary to update the hyperparameters given each additional evidence point; instead, a more efficient strategy is to update them with every $T_\text{update}$ additional points, for some specified value for $T_\text{update}$. 

Finally, from $d(\theta)$ we can retrieve a suitable pointwise approximation to the likelihood function
\begin{align}
    L(\theta) \approx \Phi\left( \frac{h - \mu_t(\theta)}{\sqrt{v_t(\theta) + \sigma^2}} \right),
\end{align}
where $\Phi(\cdot)$ is the Gaussian cumulative density, and $h$ is a threshold parameter which we choose to be the minimum of the mean discrepancy $\mu_t(\theta)$, although other choices are reasonable. The discrepancies in the evidence set can be log-transformed for possibly improving the GP-fit as advised in \citet{gutmann2016bayesian}. The rest of the steps of the algorithm remain the same. 

In practice, an acquisition function is used to determine the locations of relevant parameter space, and there are several reasonable choices for it. Some are directly based on optimizing the density fitting \citep{jarvenpaa2019} while some are efficient for finding the optimum of the function, such as the lower confidence bound selection criterion (LCBSC) \citep{srinivas2010, brochu2010}. LCBSC is defined as
\begin{align} \label{eq:lcbsc}
    \mathcal{A}_t(\theta) &= \mu_t(\theta) - \sqrt{\eta^2_t v_t(\theta)},
\end{align}
where $\eta^2_t = 2\log\left(t^{\frac{d}{2}+2}\pi^2 / 3\epsilon_\eta\right)$ is a coefficient depending on the iteration $t$, the dimension of the parameter space $d$ and the tunable parameter $\epsilon_\eta$. The parameter value is obtained by minimizing \eqref{eq:lcbsc} and varying it randomly to further balance exploration and exploitation of the probabilistic function $d(\theta)$. 
Our approach is to query the simulator at $\theta_t$, where
\begin{align}
    \theta_t \sim TN(\hat{\theta}_t, \sigma^2_\text{acq}, a_L, a_U),
\end{align}
where $TN(\hat{\theta}_t, \sigma^2_\text{acq}, a_L, a_U)$ is a normal distribution truncated to interval $[a_L,a_U]$ with mean $\hat{\theta}_t = \arg \min_\theta \mathcal{A}_t(\theta)$ and tunable variance $\sigma^2_\text{acq}$. Variance $\sigma^2_\text{acq}$ control the variation from the LCBSC minimum. The interval $[a_L, a_U]$ is the optimization region for the parameter.

BOLFI can be thought as an extension  of either an ABC- or a synthetic likelihood (SL)-type method \citep{wood2010}. 
As an SL-type method, the likelihood surrogate and the prior can be used as a target for an MCMC or a Hamiltonian Monte Carlo (HMC) sampling algorithm to draw an approximate posterior sample of size $N_\text{sample}$. A popular HMC sampling algorithm is the no-u-turn sampler (NUTS) \citep{hoffman2014} which avoids certain sensitivity issues regarding user-defined parameters which the algorithm sets automatically. In this study we use the BOLFI implementation available in ELFI, and the specific version of it is presented in Algorithm \ref{alg3:BOLFI.alg}.

\begin{algorithm}[!ht]
\caption{BOLFI}
\begin{algorithmic}
\For{$t = 1, \ldots, N_\text{init}$}
	\State Generate $\theta_t \sim {\color{black} p}(\theta)$,
    \State Generate  $y\pro \sim {\color{black} p}\left(y\mid \theta_t\right)$
 	\State Calculate $d_t = {\color{black} d}(s(y\obs), s(y\pro))$
\EndFor
\State Set $\mathcal{E}_{N_\text{init}} = \{(\theta_t, d_t)\}_{t=1}^{N_\text{init}}$
\State Fit $d(\theta) | \mathcal{E}_{N_\text{init}} \sim GP(\mu_{N_\text{init}}(\theta), v_{N_\text{init}} + \sigma^2)$
\State Set k = 0
\For{$t = N_\text{init} + 1, \ldots, N_\mathcal{E}$}
    \If{$t \equiv T_\text{update} \pmod{T_\text{update}}$}
    
    \State Optimize GP-hyperparameters \EndIf
    \State Calculate $\eta^2_t = 2 \log\left(t^{\frac{d}{2}+2 }\pi^2/3 \epsilon_\eta\right)$
 	\State Calculate $\hat{\theta}_t =  \arg \min_{\theta}\mu_t(\theta) - \sqrt{\eta^2_tv_t(\theta)}$
 	\State {\color{black} Generate} $\theta_t \sim TN(\hat{\theta}_t, \sigma^2_\text{acq}, a_L, a_U)$
 	\State Generate  $y\pro \sim {\color{black} p}\left(y\mid \theta_t\right)$
 	\State Calculate $d_t = {\color{black} d}(s(y\obs), s(y\pro))$
 	\State Set $\mathcal{E}_t = \mathcal{E}_{t-1} \cup \{(\theta_t, d_t)\}$
 	\State Fit $d(\theta) \mid \mathcal{E}_{t} \sim GP(\mu_{t}(\theta), v_{t} + \sigma^2)$ based on $\mathcal{E}_{t}$
 	\State Set k = k + 1
 \EndFor
\State Define $L(\theta) \approx \Phi\left( \frac{h - \mu_t(\theta)}{\sqrt{v_t(\theta) + \sigma^2}} \right)$
 \State Generate $N_\text{sample}$ draws from approximate posterior $\propto L(\theta){\color{black} p}(\theta)$ 
\end{algorithmic} \label{alg3:BOLFI.alg}
\end{algorithm}

\subsection{Recent progress in method development}

The features of rejection ABC, i.e.~the distance function, the distance threshold and the summarising statistics are still common in many modern ABC algorithms, even if newer methods have achieved superior performance relative to the original formulation of the method.

The basic rejection ABC uses samples generated from the prior, which can be extremely ineffective depending on the informativeness of the prior. 
ABC-PMC and ABC-sequential Monte Carlo (ABC-SMC) methods \citep{toni2009} improve on rejection ABC by sequentially constructing better proposal distributions. Sequential ABC methods have further been improved upon e.g.~by introducing adaptivity to the threshold selection \citep{simola2020adaptive} and the distance function \citep{prangle2017}. 
Other approaches have been introduced to help with summary statistic design and selection \citep{fearnhead2012, thomas2019}.
MCMC procedures can also used to generate samples from the approximate posterior distributions 
\citep{marjoram2003, marjoram2013, vihola2020}, and improving their applicability to high-dimensional problems is an on-going research problem \citep{rodrigues2019likelihood,  clarte2019component}.

Parametric approaches such as SL trade the requirement for a distance function and a threshold for a pointwise approximation of the likelihood by a parametric distribution such as a normal distribution. The moments of the approximating likelihood are, in turn, estimated using simulator draws that are possibly summarised \citep{wood2010, price2018}. The parametric approximation of the posterior enables the use of a variety of methods to draw a representative sample of the posterior \citep{an2019}. SL methods have also been extended to high-dimensional parameter spaces \citep{ong2018}.

Recently, density estimation techniques based on deep neural network architectures have been developed for  likelihood-free inference \citep{grazian2019, lueckmann2018, papamakarios2019}. These approaches are similar to the SL-type methods where the likelihood is approximated with a parametric surrogate, but in the place of the parametric distribution, neural network architectures such as autoregressive flows or emulator networks are used to fit {\color{black} either a surrogate model for the likelihood (local approach) or for the simulator (global approach)}. Furthermore, these methods have also been combined with active learning approaches to minimize the required number of simulator queries. {\color{black} \citet{lueckmann2018} proposed to minimize the variance of the posterior surrogate as suggested by \citet{jarvenpaa2019} in local inference and to maximize information gain in global inference \citep{2011arXiv1112.5745H, https://doi.org/10.48550/arxiv.1703.02910, https://doi.org/10.48550/arxiv.1710.07283}.}

\section{ABC in infectious disease epidemiology with application to Ebola outbreaks}

Simulator-based inference is well suited to infectious disease epidemiology. For example, it has been used to resolve the outbreak dynamics of stochastic birth-death-mutation models \citep{lintusaari2019}, and to infer the transmission dynamics of the Ebola haemorrhagic fever outbreak in 1995 in the Democratic Republic of Congo \citep{mckinley2009} and the COVID-19 pandemic \citep{Chinazzi395}.
In this case study, we demonstrate the application of simulator models to gain insight into the epidemic caused by the emergence of the Ebola virus in West Africa in 2014  as reported by the WHO Ebola Response Team (Team WER) \citep{who2014ebola} to infer the basic reproduction number $R_0$,  i.e.~the mean value of secondary infections caused by an infectee when no control measures are in place. 
The computational complexity of the recent, individual-based simulator models can be substantial and the pace at which inference can be delivered is of particular importance to determine the severity of the situation, and the predicted progress under different hypotheses. 
Therefore, active learning-based  methods such as BOLFI in Algorithm \ref{alg3:BOLFI.alg} provide a strategy to minimize the number of simulator queries and to quickly approximate {\color{black} the}  posterior for $R_0$.

\subsection{Team WER model}

\citet{who2014ebola} estimated the basic reproduction number by modeling the incidence of onset $I(\cdot)$ at time $t$ with a Poisson process
\begin{eqnarray}
I(t) \sim \mathsf{Poisson}\left(R_0 \sum_{s=1}^{T_\text{exp}} \omega I(t - s)\right), \quad t = 1,\ldots, T_\text{exp}
\end{eqnarray}
which is defined via the basic reproduction number $R_0$, the serial interval parameter $\omega$, which is the time difference between the onset of symptoms of the infector and the infectee, and the incidence history $I(0), \ldots, I(t-1)$.  
The serial interval parameter $\omega$ was determined heuristically from the contact tracing data. It was approximated as gamma-distributed with mean $15$ and coefficient of variation $0.66$. 
Finally, the time interval $[0, T_\text{exp}]$ is the initial period of exponential growth of infections.

In this case, the likelihood for the observed time series can be written as
\begin{eqnarray}
p(I(1), \ldots, I(T_\text{exp}) \mid R_0, \omega) = \prod_{t=1}^{T_\text{exp}} \mathsf{P}(I(t) = k \mid R_0, I(1), \ldots, I(t-1), \omega),
\end{eqnarray}
 and the posterior distribution can {\color{black} be} solved analytically when $R_0$ is modeled as gamma-distributed \textit{a priori}. However, \citet{britton2019} argued that even though this population level model has good predictive power, it would tend to lead to underestimation of $R_0$. The main causes of the underestimation in the model are the use of the serial intervals instead of generation times which indicate the time between infections of the infectee and infector, and the handling of the contact tracing when there are multiple possible infectors. 
Both of these aspects, plus other complex factors in the observation process, are simpler to take into account when using a forward simulator-based modeling approach, such as the one described in the next subsection.

\subsection{Description of the simulator}
\begin{figure}[!ht]
  \centering
  \includegraphics[trim={0 0 0 0}, clip, width = 0.9\textwidth]{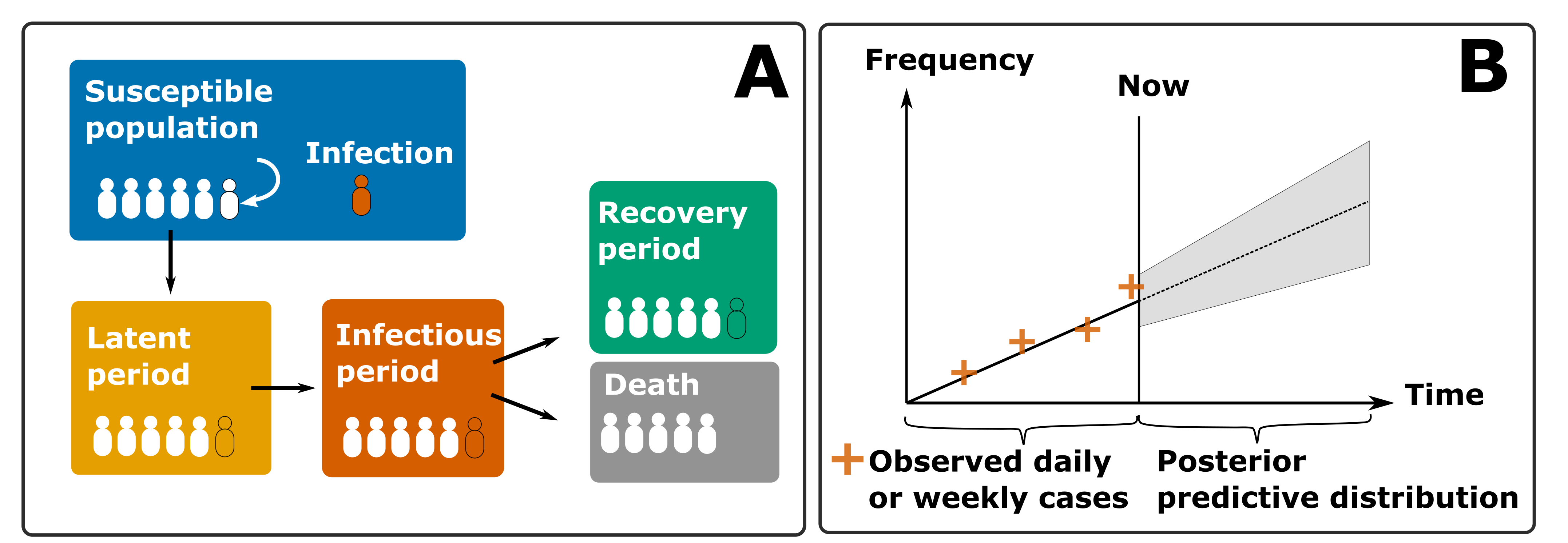}
  \caption{Panel A: In the generative model, each infected subject is in one of the four stages. After the initial infection,  the subject is in latent stage and becomes infectious after that. The infectious period ends either in recovery or in death. Panel B: The generative model can be used with the approximate posterior distribution of the basic reproduction number to probabilistically forecast how the infection will spread in the following weeks.}
  \label{Fig:ebola-diagram}
\end{figure}

We simulate an outbreak of a virus in a homogeneous and infinite community. The general simulation model follows closely the description in \citet{britton2019}. The main difference from the model used by \citet{who2014ebola}, is that the forward simulation program generates the state of each infected individual as described by the diagram in Figure \ref{Fig:ebola-diagram} instead of via a population level Poisson-process model.
Each infected individual will be in one of four possible states: \textit{latent}, \textit{infectious}, \textit{recovered} or \textit{perished}. 
After the initial infection, an individual enters the latent period $t_\text{lat}$ of the infection. It is modeled as $ t_\text{lat} \sim \Gamma(2, 5)$, where $\Gamma(\alpha, \beta)$ is a gamma-distribution with shape $\alpha$ and scale $\beta$. 
The latent period is followed by the infectious period $t_\text{inf} \sim \Gamma(1, 5)$ in which the transfer of the infection to other individuals is possible. The probability of a new infection after $s$ time units since the initial infection follows the infection rate $R_0 f_G(s)$ defined by $R_0$ and the generation time distribution $f_G(\cdot)$. The individual survives the infectious period with probability $p_\text{reco} = 0.3$ after a recovery period of $t_\text{reco} \sim \Gamma(4, 3)$, or perishes with probability (1-$p_\text{reco}$) after a period of $t_\text{die} \sim \Gamma(4/9, 9)$.

For simplicity, we assume that there is no delay in reporting the infection once symptoms arise, and that all of the cases are reported. Assumptions concerning possible reporting bias could be built into the model in a similar fashion as for the other quantities. The time period between infection and symptoms is defined as the incubation period which is the latent period multiplied by an incubation factor $\gamma \sim \unif{0.8, 1.2}$. 

The simulation always starts with a single infected individual and iterates in steps of $\Delta t = 0.2$ days until either $104$ weeks pass or $100,000$ individuals have been infected. During each time step the status of each infected individual is checked, and if the current status is \textit{infectious}, a new individual is infected with probability $p_\text{inf} = \Delta t / \Delta T$, where $\Delta T = \hat t_\text{inf} / R_0$ is the mean time between infections and $\hat t_\text{inf}$ is the mean duration of infectivity.  

{\color{black} Notice that in principle it could be possible to construct a likelihood function in this example. However, as the state of each of the infected individual is modeled separately and the number of infected individuals grows exponentially, the resulting combinatorial complexity very quickly prohibits the construction of the likelihood function outside of very small scale simulations.}

\subsection{Inference with BOLFI}

The observed data consisted of the daily cumulative count of confirmed cases from the beginning of the $2014$ epidemic for which we infer $R_0$. We used the same initial time periods $[0, T_\text{exp}]$ for estimating $R_0$ as used in \citet{who2014ebola}. 
The inference was performed separately for Guinea and Liberia using data from March $22$ -- March $30$ 2014 and June $16$ -- August $20$ 2014, respectively. Subsequent data from the following days were used to illustrate the progression of the virus. In the original study, the authors determined that the initial growth period would be over during the later time period.

The fact that we do not know the true onset of the virus is taken into account  by generating artificial counts until we obtain one that exceeds the first observed count, and then continuing the simulation for the same number of days that is in the observed dataset. This way the artificial datasets will have a similar level of variability to the observed one.

The prior for $R_0$ is modeled as a truncated normal distribution $TN(1.7, 0.5, 1.05, 4)$. As summary statistic we used the median of slopes of log-transformed case counts that were calculated with consecutive datapoints. 
As the discrepancy measure ${\color{black} d}(\cdot, \cdot)$ we used the logarithm of the euclidean distance. The BOLFI parameters are listed in Table \ref{table.BOLFI.pars.ebola}. 

To verify $R_0$ inference performance, a sample of size $500$ was drawn from the approximate posterior distribution obtained by BOLFI using NUTS sampler. The approximate posterior sample was propagated using the forward simulator, which allows for the investigation of the prediction intervals as a function of time. The propagation of the sample for the following days was visually compared to the actual observed cumulative case count.

\begin{table}[ht!]
\begin{center}
\begin{tabular}{c|c}
BOLFI-parameter & value \\\hline
$N_\text{init}$     &  \texttt{5} \\
$N_\mathcal{E}$     &  \texttt{100} \\
$T_\text{update}$ & \texttt{5} \\
$\sigma^2_\text{acq}$ & \texttt{0.1} \\
$N_\text{sample}$ & \texttt{2000}
\end{tabular}    
\end{center}
\caption{BOLFI parameters for the inference of the epidemiological model}
\label{table.BOLFI.pars.ebola}
\end{table}

\subsection{Results using empirical data}

\begin{figure}[!ht]
     \centering
     \includegraphics[trim={0 0 0 0}, clip, width = 1.0\textwidth]{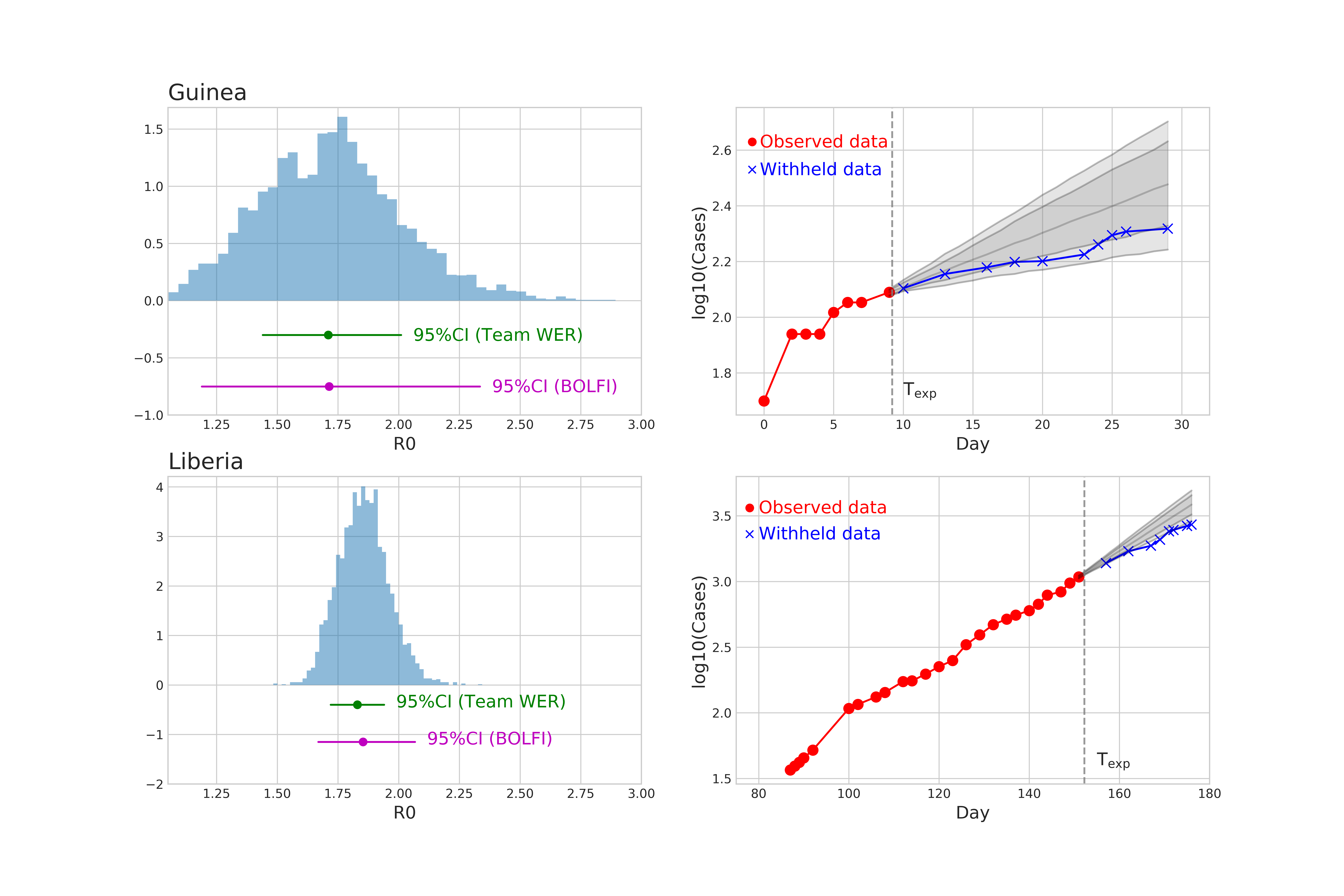}
     \caption{ Left: Approximate posterior distributions for $R_0$ using data from the initial unconstrained growth period. Results indicate that the subject-level modeling increases the uncertainty about the $R_0$ value relative to results reported in \citet{who2014ebola}. Right: The forecast of the number of cumulative infections along with observed data that were used in inference and data withheld from the inference. Ribbons on the plot indicate pointwise $80\%$ and $95\%$ probability intervals of the forecast. Comparison of forecast and withheld data indicates that period of initial growth $T_\text{exp}$ as reported in \citet{who2014ebola} is reasonable.}
     \label{Fig:ebola-posterior-predictive-guinea-liberia}
 \end{figure}

The inference results for Guinea and Liberia are illustrated in Figure \ref{Fig:ebola-posterior-predictive-guinea-liberia}. For Guinea and Liberia, the inferred estimates and $95\%$ credible intervals (CIs)   of the basic reproduction number are quite consistent with those reported by \citet{who2014ebola}. For Guinea, they reported an estimate $1.71$ with  $95\%$ CI of $1.44$--$2.01$, whereas the BOLFI posterior had a mean of $1.72$ and a $95\%$ CI of $1.19$--$2.33$. For Liberia, Team WER reported an estimate $1.83$ with $95\%$ CI, $1.72$--$1.94$ and BOLFI provided an approximate posterior with mean $1.87$ and $95\%$ CI $1.49$--$2.18$. Reference \citet{althaus2014} reported maximum likelihood estimates $1.51$ ($95\%$-confidence interval: $1.50-1.52$) and $1.59$ ($95\%$-confidence interval: $1.57-1.60$), for Guinea and Liberia, respectively. The wider uncertainty about $R_0$ in our case study CIs reflects the more complex underlying modeling assumptions, and taking such uncertainty appropriately into account may be critical in epidemic management situations.

Forward propagated posterior prediction results consisting of  the pointwise median of the simulated trajectories and the $80\%$- and $95\%$-probability bands are also reported in Figure \ref{Fig:ebola-posterior-predictive-guinea-liberia} along with the observations used in the inference and some out-of-sample observations that were not used in inference, but are used to illustrate forecasting accuracy.
\citet{who2014ebola} assumed, on a basis of a visual inspection, that the initial period of exponential growth would be $30$ March for Guinea and $24$ August for Liberia which are illustrated with the gray vertical lines on Figure \ref{Fig:ebola-posterior-predictive-guinea-liberia}. Dates are reported as counts since the beginning of the epidemic (March $22$). 
Our model fit is consistent with the initial growth period assumption as the observed data counts after the assumed initial growth period diverge quickly from the probable prediction curves.
 
\section{ABC in personalised cancer treatment with application to breast tumor modeling}

Personalised cancer treatment is an application area where simulator-based inference has a lot of potential, given the constantly improving biological generative models for the evolution of the disease under treatment \citep{sottoriva2010, kozlowska2018, lai2019, Lai2021}. 
Realistic biological generative tumor models that are built up from the molecular and cellular processes result typically in a level of complexity that renders analytical solutions infeasible.
Using simulator-based methods we are able to conduct inference on such detailed models, which could in the future be used to optimize personalised treatments to achieve the best possible therapy outcomes.
However, the complexity of the biological modeling often makes the generative modeling computationally demanding and only active learning based methods such as BOLFI (Algorithm \ref{alg3:BOLFI.alg}) enable the inference to be performed within a reasonable time.

\subsection{Description of the Simulator}

We consider an example of breast cancer tumor modeling that is a combination of empirical data and detailed computer simulation. The system is partly initialised and tuned based on a biopsy from a real patient, but we use only simulated data to demonstrate the inference and prediction procedure.

\begin{figure}[htbp]
    \centering
 \includegraphics[width = 0.9\textwidth]{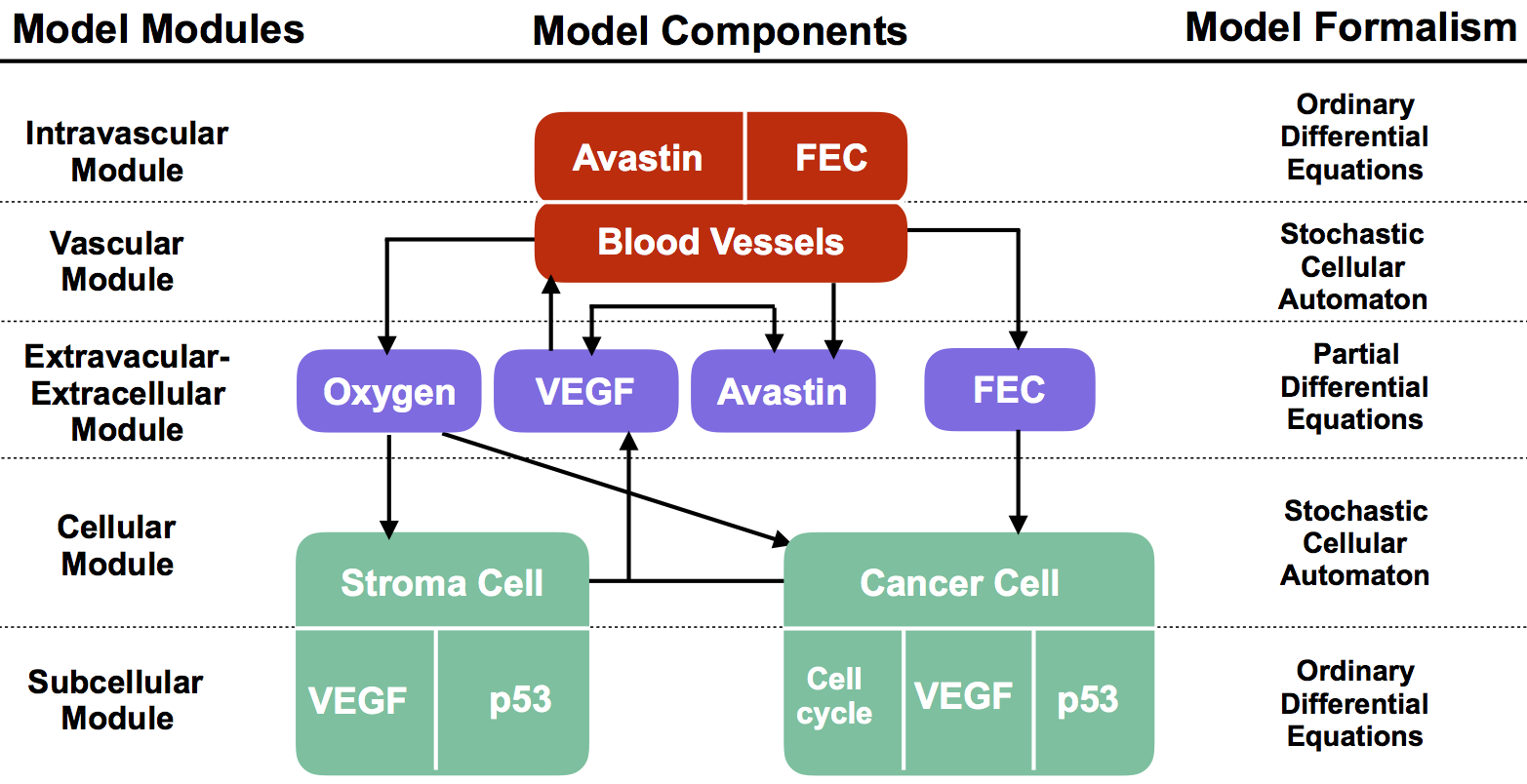} 
    \caption{Modular structure of the hybrid cellular automaton describing the response of breast tumor tissue to a cocktail of chemoterapies (FEC) and Avastin. The diagram shows the main components of each module and the interactions among them. The right column shows the different model formalisms used for each of the model modules. See ~\citep{lai2019} for full model details and its patient-specific parameterisation.  
 }
    \label{Fig:cancer_model}
 \end{figure}

The cancer simulator we use is a multi-scale pharmacokinetic and pharmacodynamic model describing the response of a cross-section of breast tumor tissue to a combination of chemotheraupetic and anti-angiogenic agents~\citep{lai2019}.  Mathematically, it consists of a  hybrid cellular automaton model  \citep{Ribba2004, alarcon2010} that couples stochastic and discrete model formalisms with deterministic and continuous components accounting for biological processes at different spatio-temporal scales; see Figure~\ref{Fig:cancer_model}.
There are multiple parameters in the model influencing the evolution of cancer cells, stromal cells and tumour vessels, which are all modeled as individual agents. 
Many of the parameters can be inferred and fixed via various means but some of the unknown parameters have to be inferred based on the reaction of the tumour to the treatment. 
Here we focus on the inference of two key parameters determining the outcome of each patient to the drug treatment. Those parameters account for the  \textit{sensitivity of cancer cells to the chemotherapy}, $\logcs$, and the \textit{minimal cell cycle length of cancer cells}, $\tcell$.
We can write the simulator as a nonlinear time series model for the evolution of state $x_t$
\begin{align}
    x_{t} = f_{t}(\logcs, \tcell, x_{t-1},  v_{t}), \quad t = 1, 2, \ldots, T,
\end{align}
where $f_{t}$ is the nonlinear transition model at $t$, which is the time for the temporally discretized simulator with $30$ minute increments, and $v_{t}$ is the stochastic component of the simulator.  
The simulated state $x_{t}$ consists of  cells, vessels, and extracellular concentrations of oxygen, Avastin and vascular endothelial growth factor (VEGF) within the simulation grid, which here is a $33 \times 20$ rectangular grid representing a specific two-dimensional cross-section of the tumor.

The time series model is not available for constant monitoring but some of the components can be indirectly measured at sparse time points using methods such as magnetic resonance imaging. For simplicity we assume that it is possible to measure the true state of the system
\begin{align}
    y_{t_k} &= x_{t_k}, \quad k = 1, 2, \ldots ,
\end{align}
In this proof-of-concept study we assume that observations can be collected every three days $t_k = k \cdot 3 \cdot 48, ~ k = 1, 2, \ldots, 6$. 

The drug is administered for the patient every $3$ weeks, i.e.~at times $t^\text{drug}_k = k \cdot 21 \cdot 48, ~  k = 0, 1, 2, 3$ and we aim to infer the parameters within the first $3$-week treatment period, as we would want to forecast whether or not the chosen treatment is effective before the second dose. This dose could be adjusted, given the inferred parameters, to achieve an optimal outcome.
Being able to reliably infer the parameters using as little data as possible from the beginning of the cancer evolution curve would enable simulated testing of personalised treatment strategies that are anticipated to be efficient for the specific patient. 

\subsection{Inference with BOLFI}

We generate the fake observed data using fixed parameter values $\tcell = 14.69$ and $\logcs=3.0$, and predetermined treatment protocol, and investigate how well we can infer the set parameters and how accurately we are able to forecast the disease progress based on the data collected within the first treatment period.
If we were able to infer the patient specific parameter values based on the observed data, then we would be able to forecast how the patient defined by the parameter values would react also to \textit{different} treatment protocols within the accuracy of the simulator. 
In our inference framework the results are approximations to the posterior distribution of the parameters $p(\logcs, \tcell \mid y_{t_1}, \ldots y_{t_6})$, conditioned on the state observed at $t_1, \ldots t_6$.
Prior distributions for the unknown parameters are modelled as uniform $\logcs \sim \unif{1,4}$ and $\tcell \sim \unif{1,21}$

As summary statistics we use the proportions of cancerous cells in the simulation grid at each observation time,
\begin{align}
s_k := s(y_{t_k}) &= \frac{1}{660}\sum_{i = 1}^{660} y_{t_k,\text{cells}}^{(i)}, \quad k = 1, \ldots, 6
\end{align}
where $660 = 33 \cdot 20$, is the size of the simulation grid and $y_{t,\text{cells}}^{(i)}, i = 1, \ldots, 660$ is the vectorized simulation grid with binary value $1$ when the $i$th grid location contains a cancer cell.
The simulations are computationally very demanding and we use BOLFI to produce the approximation of the posterior. The BOLFI parameters are listed in Table \ref{table.BOLFI.pars.cancer}. We use the log-transformed euclidean distance for the response of the surrogate model within BOLFI.

\begin{table}[htbp]
\begin{center}
\begin{tabular}{c|c}
BOLFI-parameter & value \\\hline
$N_\text{init}$     &  \texttt{30} \\
$N_\mathcal{E}$     &  \texttt{150} \\
$T_\text{update}$ & \texttt{10} \\
$\sigma^2_\text{acq}$ & \texttt{[0.5, 0.1]} \\
$N_\text{sample}$ & \texttt{2000}
\end{tabular}    
\end{center}
\caption{BOLFI parameters for the inference of the breast cancer model}
\label{table.BOLFI.pars.cancer}
\end{table}

After obtaining an approximate posterior curve based on the likelihood surrogate provided by BOLFI, NUTS is used to draw a sample from it, and the generative model is used to propagate the posterior sample in time under the selected treatment. This results in a simulation-based estimate of the posterior predictive distribution of the state $p(x_{t_\text{end}} \mid s_1, \ldots s_6)$. Here we use the twelve week mark $t_\text{end} = 4032$ as the end point. This time interval contains four full treatment periods.

\subsection{Results}\label{sec:cancessimresults}

\begin{figure}[!ht]
    \centering
 \includegraphics[width = \textwidth]{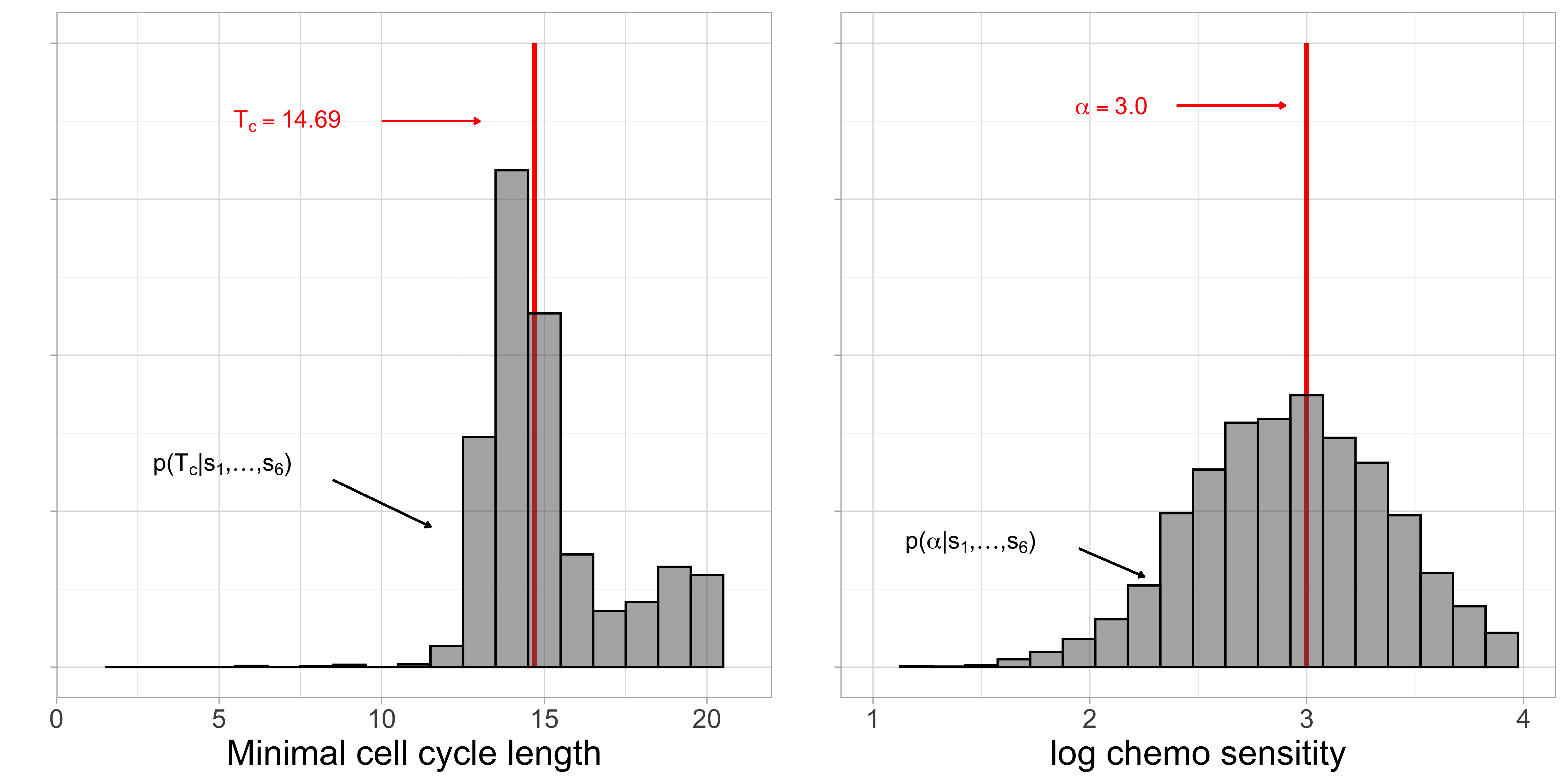} 
    \caption{ Approximated posterior distributions of the posterior marginals. The red vertical lines on the first two subplots indicate the true simulation values which were used for simulating the cancer cell growth trajectory.} 
    \label{Fig:cs_distributions}
 \end{figure}

\begin{figure}[!ht]
    \centering
 \includegraphics[width = \textwidth]{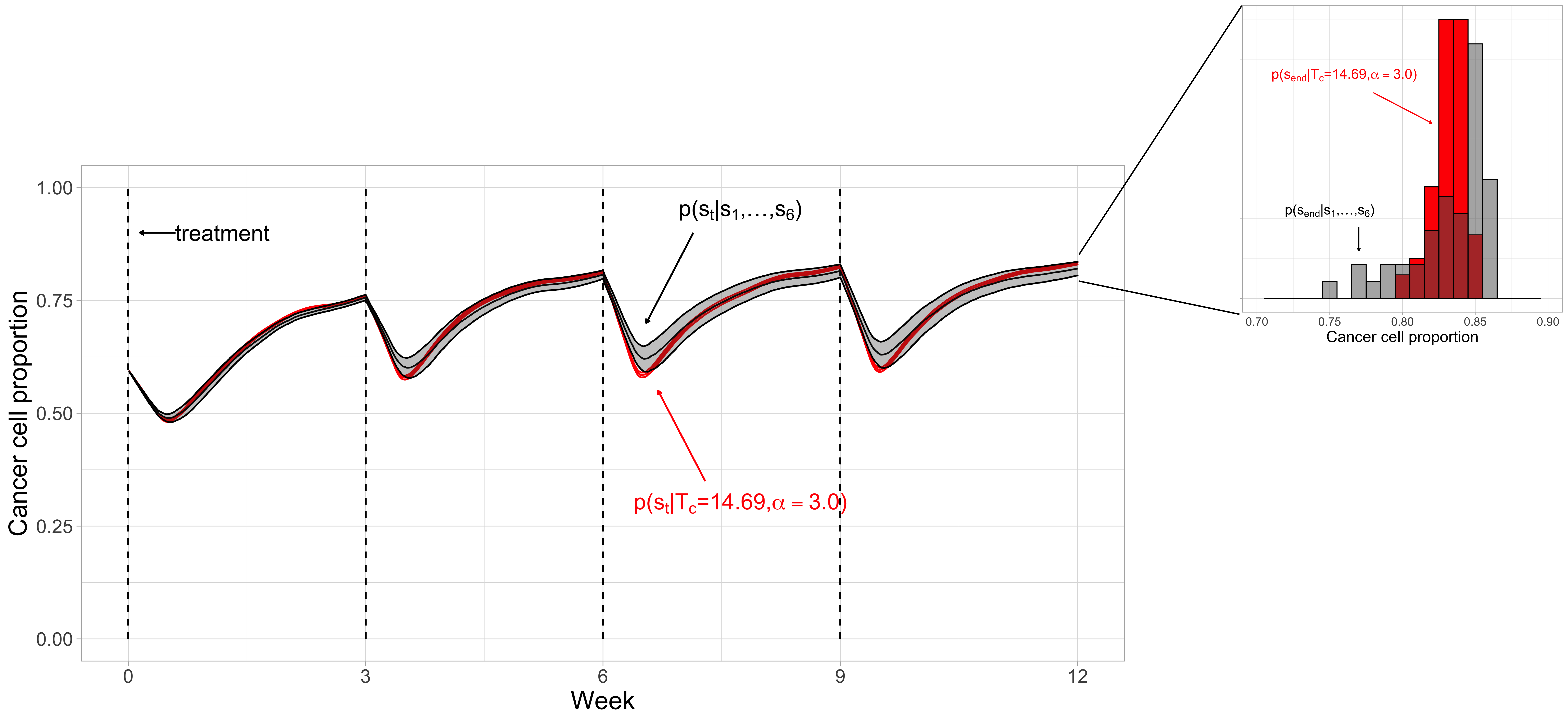} 
    \caption{Simulated trajectories of the evolution of cancer cell proportions in the simulation grid given the true parameter values, and the forecast of the trajectory as simulated given the posterior distribution of the parameter values. The simulation end point is highlighted as the histogram on the right. \label{Fig:cs_forecast}} 
 \end{figure}

The inference results are illustrated in Figure \ref{Fig:cs_distributions}. We see that the posterior distributions have the probability mass peaks located closely around the true values. Importantly, we are able to use the posterior distributions to probabilistically investigate how the disease will evolve under treatment. In Figure \ref{Fig:cs_forecast} we plot the summarised state of the cancer cells for $12$ week treatment period and compare it to the trajectory simulated given the true parameter values. The results indicate that the forecast is quite consistent given the inferred posterior distributions and that the current treatment most likely will not erase the cancerous growth, and other treatments should be investigated for a better outcome.

%


\section{ABC in astronomy with an application to supernova models} \label{sec.intro}

Likelihood--free inference is becoming increasingly important in astronomy,
 where physical models cannot often be fully characterized in terms of a tractable likelihood function \citep{ShaferFreeman2012, CameronPettitt2012, WeyantEtAl2013, IshidaEtAl2015, leclercq2018bayesian, picchini2020adaptive}. Here we evaluate the performance improvement arising from using BOLFI (Algorithm \ref{alg3:BOLFI.alg}) instead of the ABC--PMC algorithm (introduced in  Section  \ref{intro.abcpmc}) on an astronomical model from \cite{jennings2017astroabc}. The ABC--PMC version proposed for this example is slightly different from Algorithm \ref{alg2:ABC--PMC.alg}. In particular, the version employed for this example requires us to properly tune the final number of iterations $T$, the first tolerance $\epsilon_1$ and the quantile $q_t$ used for adaptively decreasing the series of tolerances. We did so by following the suggestions in \cite{Lenormand2013} and \cite{WeyantEtAl2013}. The choices for these three quantities are highlighted in Section \ref{sec.methods}.

Starting with the SuperNova ANAlysis (SNANA) light curve package by \cite{kessler2009snana} and the corresponding implementation of the SALT--II light curve fitter presented in \cite{guy2010supernova}, a sample of $400$ supernovae with redshift range $z \in [0.5, 1.0]$ have been simulated and then binned into $i = 20$ redshift bins. A model that describes the distance modulus as a function of redshift $z$ is defined as:  
\begin{equation}
\mu_{i}^\text{model}(z_{i};\Omega_{m},w_{0}) \propto 5\log_{10}\biggl(\frac{c(1+z_i)}{h_{0}}\biggr)\int_{0}^{z_i}\frac{dz'}{E(z')},
\label{eq:supernovaFM}
\end{equation}
where $E(z) = \sqrt{\Omega_{m} (1+z)^{3} + (1-\Omega_{m})\exp\left(3\int_{0}^{z} \text{d}\ln(1+z')[1+w(z')]\right)}$.

The true cosmological parameters used to generate the ``observed'' data $\mu$ are the matter density of the universe $\Omega_m = 0.3$, the dark energy density of the universe $\Omega_{\Lambda} = (1 - \Omega_m) = 0.7$, the present value of the dark energy equation $\omega_0 = -1.0$ and, finally, the current Hubble constant $h_0 = 0.7$. In the following, $\Omega_{\Lambda}$ and $h_0$ are considered known and fixed at their input values. The goal is to estimate the two cosmological parameters $\Omega_m$ and $\omega_0$. The original example as presented in \cite{jennings2017astroabc} is available in the \textit{astroABC} Python package. 

\citet{jennings2017astroabc} added artificial noise to the data simulated through \eqref{eq:supernovaFM} by using a skew--normal distribution \citep{azzalini1985class} with location, scale and skewness parameters fixed at $ -0.1 $, $0.3$ and $5.0$, respectively. 
By doing so, the commonly used MCMC algorithm for Bayesian statistical inference is not applicable. In fact, in order the perform the analyses by using the MCMC algorithm, \cite{jennings2017astroabc} tried to add artificial noise to the data simulated through Eq. \eqref{eq:supernovaFM} by using a normal distribution with location and scale parameters fixed at $ -0.1 $ and $0.3$, respectively. The results obtained by the MCMC algorithm (see Section 5 in \cite{jennings2017astroabc}) led to a poor estimation of the parameters of interest  and therefore an ABC based approach is preferred \citep{jennings2017astroabc}.

The goal of the analysis presented in \cite{jennings2017astroabc} was to present a comparison between this slightly more complicated model (for which the ABC--PMC algorithm is required) and a simplified version for which artificial noise is not added (for which the likelihood function is tractable and therefore MCMC is possible). The contour plot of the joint distribution ($\Omega_m$, $\omega_0$)  obtained by using the MCMC algorithm and the marginal posterior means for $\Omega_m$ and $\omega_0$ are available in \cite{jennings2017astroabc}. Beyond retrieving reliable summaries such as marginal posterior means and marginal highest posterior density (HPD) intervals for the parameters of interest, it is of relevance from a physics standpoint to evaluate how well a likelihood-free inference approach preserves the so--called ``banana--shape'' \citep{kessler2013testing, hinton2019steve} that describes the relation between $\Omega_m$ and $\omega_0$.  The ``banana--shape''  is not expected to significantly change after the artificial noise is added to the data, as shown by \cite{jennings2017astroabc}.

In order to use BOLFI (introduced in Section \ref{intro.bolfi}) and the ABC--PMC sampler for the estimation of the matter density of the universe parameter $\Omega_m$ and the dark energy equation parameter $\omega_0$, two additional quantities must be specified. As highlighted in Section \ref{sec.stat.met}, the distance function used to compare the  observed  and the  simulated data and the prior distributions for the parameters of interest (i.e. for this example $\Omega_m$ and $\omega_0$) must be defined. Since in this section we want to compare the performance of BOLFI with the ABC--PMC sampler, we performed the analysis for both methods by using the same specifications recommended in \cite{jennings2017astroabc}. The metric ${\color{black} d(\cdot, \cdot)}$ that compares the observed data $\mu$ with the simulated data $\mu_\text{sim}(z)$ is defined as:
\begin{equation}
{\color{black} d}(\mu,\mu_\text{sim}(z)) = \sum_{i} \frac{(\mu_{i} - \mu_\text{sim}(z_{i}))^{2}}{2 \sigma_{i}^{2}},
\label{eq:supernova_distance}
\end{equation}
where  $\sigma_{i}$ is the error on the data point $\mu_{i}$, estimated by calculating the sample variance of the observation in the $i^{th}$ bin. We note that, in this example, \cite{jennings2017astroabc} do not formally define a summary statistic for the data. Therefore the summary statistic is equivalent to the $20$ dimensional vector of the data. The prior distributions, $\Omega_m \sim N(0.3, 0.5)$ and $\omega_0 \sim N(-1.0, 0.5)$, are chosen, where the prior for $\Omega_m$ is consistent with the $(0,1)$ range for this parameter.


 \subsection{ABC-PMC inference and acceleration by BOLFI} \label{sec.methods}

The ABC--PMC sampler from the \textit{astroABC} package was run using $N = 1000$ particles, a total number of iterations $T = 20$ and a quantile equal to $q_t = 0.75$, which is used to reduce the ABC tolerance parameter through the iterations. It follows that, with respect to the ABC--PMC sampler defined in Algorithm  \ref{alg2:ABC--PMC.alg}, the vector of the ABC tolerances $\epsilon_1, \dots, \epsilon_{20}$ is not tuned in advance by the researcher but instead defined through the approach suggested by \cite{Lenormand2013}, among others. The perturbation kernel used from the second iteration onwards in Algorithm~\ref{alg2:ABC--PMC.alg} follows the recommended Gaussian distribution, having variance equal to twice the empirical coverage amongst the particles for both $\Omega_m$ and $\omega_0$. Following the choices by \cite{jennings2017astroabc}, the first tolerance was fixed to $\epsilon_1=500$ and the final tolerance was $\epsilon_{20}=29.82$. We tried different combinations of $T, q_t$ in order to identify a suitable level for the final tolerance at which to stop the ABC--PMC algorithm, resulting in $\epsilon_T$ close to $30$.

 \begin{table}[ht!]
 \begin{center}
\begin{tabular}{c|c}
BOLFI-parameter & value \\\hline
$N_\text{init}$     &  \texttt{50} \\
$N_\mathcal{E}$     &  \texttt{300} \\
$T_\text{update}$ & \texttt{1} \\
$\sigma^2_\text{acq}$ & \texttt{1} \\
$N_\text{sample}$ & \texttt{1000}
\end{tabular}
\end{center}
\caption{BOLFI parameters for the estimation of the supernova model}
\label{table.BOLFI.pars.3}
\end{table}

\begin{figure}[!ht]
   \centering
   \includegraphics[width = 0.8\textwidth]{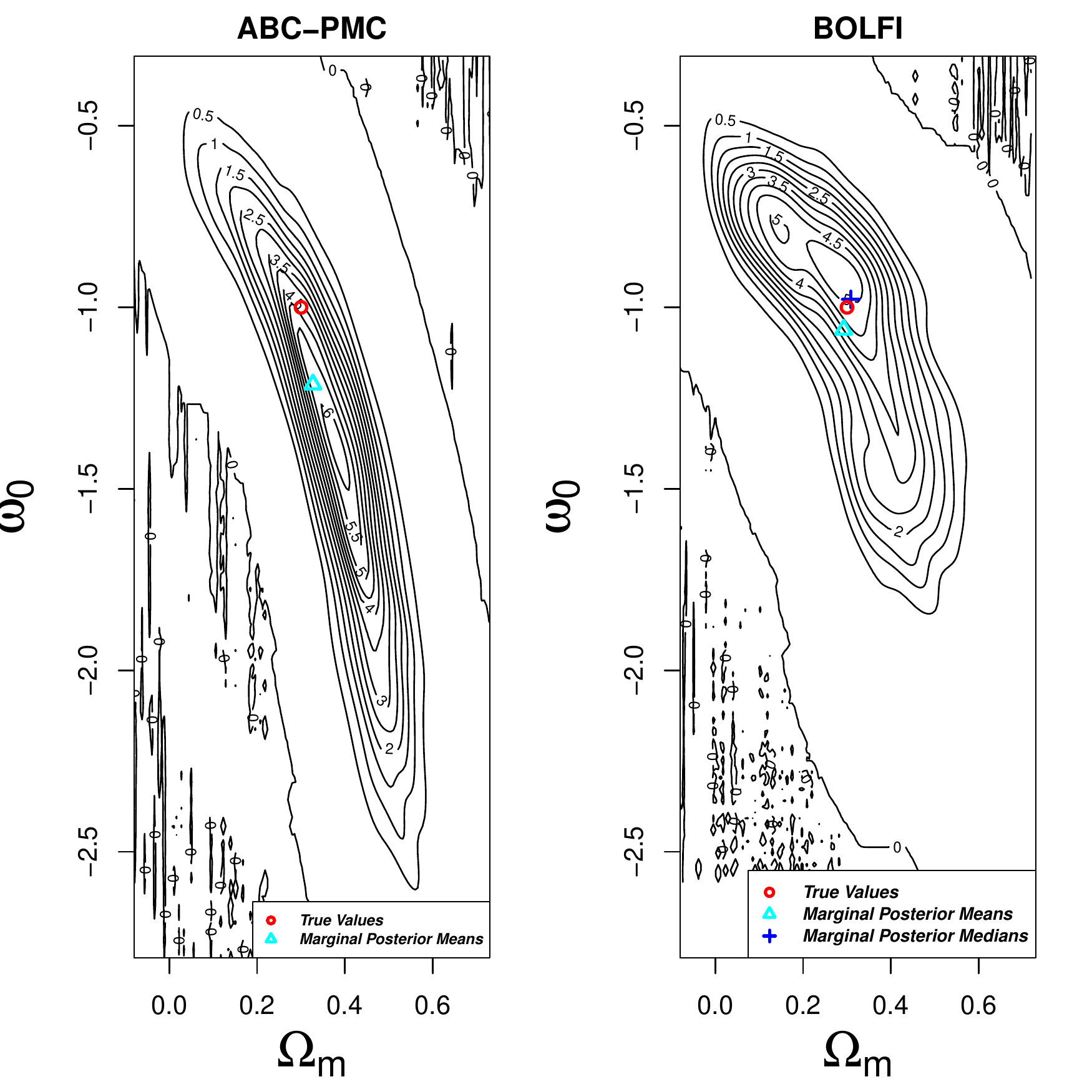} 
   \caption{(left) Contour plot of the joint distribution ($\Omega_m$, $\omega_0$) obtained by the ABC--PMC sampler and (right) contour plot of the joint distribution ($\Omega_m$, $\omega_0$) obtained by BOLFI. The true values ($\Omega_m=0.3$, $\omega_0=1$) are highlighted with a red circle. Relevant regions of the parameter space are inferred and for both methods the marginal posterior means (and for BOLFI also the marginal posterior medians) are highlighted (see also Table \ref{table.BOLFI.results}). The point estimates obtained by BOLFI are closer to the true values for $\Omega_m$ and $\omega_0$ compared with the corresponding estimates provided by ABC--PMC. The expected ``banana--shape'' is reconstructed by both methods, although the contour plot obtained by BOLFI presents a smaller lower tail compared with the ``banana--shape'' retrieved by using the ABC--PMC algorithm.}
   \label{Fig:bolfi.supernovae}
\end{figure}

The computational efficiency of both the ABC--PMC sampler and BOLFI was investigated. Our parameter choices for BOLFI are summarized in Table \ref{table.BOLFI.pars.3}. We used a Metropolis-Hasting algorithm to produce the $N_\text{sample}$ approximate posterior draws. With the selected particle sample size of $N=1000$, the ABC--PMC sampler takes $90$ minutes to produce the final ABC posterior distribution. In comparison, BOLFI produces the posterior distribution in $3$ minutes. The gain in computational efficiency is {\color{black}{a clear advantage}} obtained by using BOLFI over the ABC--PMC sampler. Figure \ref{Fig:bolfi.supernovae} displays the contour plots of the joint distribution ($\Omega_m$, $\omega_0$) obtained by the ABC--PMC sampler and by BOLFI, while the point estimates for $\Omega_m$ and $\omega_0$ obtained by the ABC--PMC analysis (the weighted marginal posterior means) and the estimates retrieved by BOLFI (the marginal posterior means and marginal posterior medians) are summarized in Table \ref{table.BOLFI.results}. It is possible to note that BOLFI provides marginal posterior means closer to the true values ($\Omega_m=0.29$ and $\omega_0=-1.06$) compared with the corresponding estimates provided by the ABC--PMC ($\Omega_m=0.36$ and $\omega_0=-1.22$). Both procedures are able to reconstruct the expected ``banana--shape'', although the contour plot obtained by BOLFI presents a smaller lower tail compared with the ``banana--shape'' retrieved by using the ABC--PMC algorithm. This observation is also confirmed by looking at the marginal $90\%$ HPD credible intervals for $\Omega_m$ and $\omega_0$, reported in parentheses in Table \ref{table.BOLFI.results}. {\color{black} However, repeated experiments would be required to quantify how well the methods based on different approximations actually estimate the posterior distributions.}

\begin{center}
\begin{table}[ht]
\centering
\resizebox{\columnwidth}{!}{%
\begin{tabular}{|c|c|c|c|c|}
\hline 
& True values & ABC--PMC ($90\%$ HPD) & BOLFI mean ($90\%$ HPD) & BOLFI median ($90\%$ HPD) \tabularnewline
\hline 
\hline 
$\Omega_m$ & $0.3$ &  $0.36$ $(0.18, 0.54)$   & $0.29$ $(0.055, 0.48)$  & $0.31$ $(0.055, 0.48)$  \tabularnewline
\hline 
$\omega_0$ & $-1$ & $-1.22$ $(-2.11, -0.62)$  & $-1.06$ $(-1.56,-0.61)$  & $-0.98$ $(-1.56,-0.61)$  \tabularnewline
\hline
\end{tabular}
}

\caption{Results obtained by using the ABC--PMC sampler and BOLFI. Together with the obtained point estimates of $\Omega_m$ and $\omega_0$ (the weighted marginal posterior means for the ABC--PMC analyses and the marginal posterior means and marginal posterior medians for BOLFI), the marginal HPD credible intervals $90\%$ are displayed. BOLFI provides point estimates for the parameters that are closer to the true values compared with the corresponding estimates retrieved by the ABC--PMC sampler.  HPD credible intervals $90\%$ for $\Omega_m$ and $\omega_0$ indicate that the probability mass in the ``banana--shape'' produced by BOLFI is within a more compact region with  with a smaller lower tail compared with the ``banana--shape'' retrieved by using the ABC--PMC algorithm.}
\label{table.BOLFI.results}
\end{table}
\end{center}

\section{ABC forecasting with an application to optimal portfolio allocation}\label{sec.forecasting}

Thus far the discussion has centred primarily on the use of ABC as an inferential
method, and on improving the performance of more basic versions of ABC via BOLFI. Whilst BOLFI has been used for \textit{prediction} in two of the three previous illustrations, any comparison with predictions that \textit{would} have been produced via exact (likelihood-based) Bayesian inference has not been possible, due simply to the fact that the likelihood function is inaccessible (or, at the very least, challenging) in the given examples.

In this section, we step back from the illustration of ABC in situations where it is essential, to an artificial situation in which the exact posterior and, hence, the exact predictive, is available. The aim of the exercise is to illustrate that an ABC-based predictive distribution can be a very accurate approximation of the exact predictive and, hence, yield equally reliable forecasts. This then provides some reassurance that prediction based on {\color{black}an LFI method} has value in those cases where it is indeed the only option, such as those illustrated in this paper. We revert here to the simplest form of rejection ABC in order to emphasize that predictive accuracy does not {\color{black} necessarily} depend on using an optimal version of the inferential algorithm.  

Let $Y_{n+1}$ denote a scalar random variable, observed at time $n+1$, and
generated according to $p(y_{n+1}|\theta,y\obs)$, where $y\obs%
=[y_{1},y_{2},...,y_{n}]{\transp}.$ The exact Bayesian predictive (or forecast
distribution - we use the
terms `forecast' and `prediction', and their variants, synonymously) is $p(y_{n+1}|y\obs)=\int p(y_{n+1}|\theta,y\obs)p(\theta|y\obs%
)\text{d}\theta\mathbf{,}$ where $p(\theta|y\obs)$ is the exact posterior, defined in the usual way,
and $y_{n+1}$ denotes a value in the support of $Y_{n+1}$. In cases
where $p(y\obs|\theta)$ and, hence, $p(\theta|y\obs)$, is
inaccessible, $p(y_{n+1}|y\obs)$ is also inaccessible, and a natural
solution is to define the \textit{approximate }Bayesian predictive,
$g(y_{n+1}|y\obs)=\int_{\Theta}p(y_{n+1}|\theta,y\obs)\pi_{\epsilon}(\theta \mid y\obs)\text{d}\theta\mathbf{,}$ with $g(y_{n+1}|y\obs)$ produced using the ABC posterior in (\ref{ABCposterior}), $\pi_{\epsilon}(\theta \mid y\obs)$,  which could be accessed using either Algorithm \ref{alg1:rejABC.alg} or Algorithm \ref{alg2:ABC--PMC.alg}. Alternatively, BOLFI (Algorithm \ref{alg3:BOLFI.alg})  could be used to produce an approximate posterior sample, as described in Section 2.3, noting that a notational change would be required to represent this approximate posterior in the expression for $g(y_{n+1}|y\obs)$. 

In an extensive exploration, \cite{frazier2019approximate}
demonstrate, both {theoretically} and in practical {situations}, that the
differences between $g(y_{n+1}|y\obs)$ and $p(y_{n+1}|y\obs)$ {can
be} \textit{negligible}, despite there sometimes being \textit{substantial }differences
between the approximate and exact posteriors. Moreover, {the authors }also
demonstrate that ABC-based forecasting can produce reliable forecasts in a fraction of the time
required for exact methods, owing to the speed with which $\pi_{\epsilon}(\theta \mid y\obs)$ can be constructed, {relative to} $p(\theta
|y\obs)$.

The simplicity and computational speed of ABC-based forecasting is important in the sphere of
economics and finance, where the need to predict the actions, and
interactions, of large numbers of economic `agents' leads to complex dynamic
models that often challenge the MCMC toolkit and exact Bayesian
forecasting.  To illustrate this novel use of ABC, we document the performance of
ABC-based forecasting, relative to exact forecasting, in a particular empirical example: the
production of a utility-optimizing financial portfolio. { We reiterate that our illustration is based on the simplest version of ABC, as given in Algorithm \ref{alg1:rejABC.alg}.}

\subsection{Optimal Portfolio Allocation: the Role of Prediction\label{opt}}

At time $n$, an investor chooses to allocate her wealth $W_{n}$ across $m$
possible investment choices, {where at time} $(n+1)$ the different investment
choices yield random returns denoted by $R_{n+1,i}$, $i=1,2,...,m$, with
$R_{n+1}=[R_{n+1,1},...,R_{n+1,m}]{\transp}$. For $\Delta_{m}%
:=\{{\alpha}\in\lbrack0,1]^{m}:\sum_{i=1}^{m}\alpha_{i}=1\}$, the
individual's wealth at time $(n+1)$ is given by $W_{n+1}=W_{n}%
[1+{\alpha}{\transp}R_{n+1}]$. The goal of portfolio analysis is to
discern an \textquotedblleft optimal\textquotedblright\ allocation rule
${\alpha}^\text{opt}\in\Delta_{m}$ for the portfolio ${\alpha
}{\transp}R_{n+1}$.

The portfolio allocation problem exhibits different solutions depending on the
definition of \textquotedblleft optimality\textquotedblright. A common
approach in {economics and finance} is to find ${\alpha}^\text{opt}$ by
maximizing expected utility (of wealth) using von Neumann-Morgenstern expected
utility (EU) theory \citep{vNM1953}: For $u(\cdot
):[0,\infty)\rightarrow\mathbb{R}$ a utility function, {and $\mathbb{E}%
_{n}(\cdot)$ denoting expectation conditional} on information available at
time $n$, the optimal allocation rule is
\begin{equation}
{\alpha}_{n+1}^\text{opt}=\arg\max_{{\alpha}_{n+1}\in
\Delta_{m}}\mathbb{E}_{n}[u(W_{n+1})]. \label{EU}%
\end{equation}

For this illustration we consider the canonical risk-averse investor with
power utility function, $u(W_{n+1})=\frac{W_{n+1}^{1-\gamma}}{1-\gamma}$, where $\gamma>1$ denotes the risk aversion parameter. Given a model for
conditional returns, $p(R_{n+1}|R\obs)$, where $R\obs =[R_{1},R_{2},...,R_{n}]{\transp}$, the allocation rule that
maximizes the expected utility is given by
\begin{align}
{\alpha}_{n+1}^\text{opt} & =\arg\max_{{\alpha}_{n+1}\in
\Delta_{m}}E_{n}\left(  u(W_{n+1})\right)   =\arg\max_{{\alpha
}_{n+1}\in\Delta_{m}}\int u(W_{n+1})p(R_{n+1}|R\obs)\text{d}R_{n+1}. \label{neweu}%
\end{align}

Given that $p(R_{n+1}|R\obs)$ is typically unavailable in closed-form, we
must resort to simulation to compute the integral in \eqref{neweu}: if we can
obtain $M$ {draws} from $p(R_{n+1}|R\obs)$, denoted as $R_{n+1}^{(j)}$,
$j=1,...,M$, the value of $\mathbf{\alpha}_{n+1}^\text{opt}$ can be approximated
numerically by solving
\begin{equation}
\hat{{\alpha}}_{n+1}^\text{opt}\equiv\arg\max_{{\alpha}%
_{n+1}\in\Delta_{m}}\sum_{j=1}^{M}u\left(  W_{n}\left[  1+{\alpha
}_{n+1}{\transp}R_{n+1}^{(j)}\right]  \right)  /M.\label{alp_opt}%
\end{equation}
The optimally allocated portfolio at time $n+1$ is then given by
$W_{n+1}^\text{opt}=W_{n}[1+\widehat{{\alpha}}_{n+1}^{\text{opt}\transp}%
R_{n+1}^\text{obs}]$, with utility, $u(W_{n+1}^\text{opt})$, where $R_{n+1}^\text{obs}$
denotes the observed value of $R_{n+1}$. Repeating this exercise over an
evaluation period yields a series of such utility values, which can be
averaged to produce an estimate of $\mathbb{E}_{n}[u(W_{n+1}^\text{opt})]$,
$\widehat{\mathbb{E}}_{n}[u(W_{n+1}^\text{opt})].$ In what follows, we demonstrate
that, in this particular representative example, there is negligible
difference between { the values of $\widehat{\mathbb{E}}_{n}[u(W_{n+1}^\text{opt})]$ produced via}
\textit{exact} and \textit{approximate} predictives associated with a given
model {for }$R_{n+1}$.

\subsection{Empirical Example\label{emp}}

For the illustration we consider a very simple portfolio, $W_{n+1}%
=W_{n}[1+{\alpha}_{n+1}{\transp}R_{n+1}],$ where $R_{n+1}%
=[\exp(y_{n+1}),\exp(rf_{n+1})]{\transp}$ and ${\alpha}%
_{n+1}=[\alpha_{n+1},1-\alpha_{n+1}]{\transp},$ with $y_{n+1}$ the
logarithmic return on the S\&P500 market portfolio and $rf_{n+1}$ the (known)
logarithmic return on a one-month constant maturity US Treasury Bill, over the period
$n$ to $(n+1).$ The monthly S\&P500 Total Returns index is sourced from the Chicago Board of Options Exchange (CBOE), and captures both
capital gains as well as dividend yields. The Treasury Bill is sourced from the Federal Reserve St Louis (FRED)
database. {The weight }$\alpha_{n+1}\in\lbrack0,1]$ determines the
proportion of wealth allocated to the risky market portfolio, and is to be chosen according to (\ref{alp_opt}). (See, for example, \citep{BCRvD2013}.) The
monthly data extends from June 1986 to June 2018, totaling 384
observations. The first 324 observations are reserved for inference/training
and the final 60 observations (five years) used to estimate expected utility.

We produce the 60 one-step-ahead exact and approximate predictive
distributions over the evaluation period based on the following stochastic
volatility model for $y_{t}$,%
\begin{align}
y_{t}  &  =\mu+\sqrt{V_{t}}\varepsilon_{t}\label{1}\\
\ln V_{t}  &  =\omega+\rho\ln V_{t-1}+\sigma v_{t}, \label{2}%
\end{align}
where $\varepsilon_{t} \overset{\text{i.i.d.}}{\sim} N(0,1)$ and $v_{t}\overset{\text{i.i.d.}}{\sim}N(0,1)$, with
$\varepsilon_{t}$ and $v_{t}$ independent for all $t = 1,2,...,n.$ The unknowns comprise the static parameter vector, $\theta=[\rho,\sigma,\omega,\mu]{\transp}$, the
vector of (in-sample) latent volatilities, ${V}=[V_{1},V_{2}%
,...,V_{n}]{\transp}$, and the unknown $V_{n+1}$ on which $y_{n+1}$ is
conditioned. This model is adequate for capturing the behaviour of monthly returns, and allows for a ready application of an exact algorithm, for this comparative exercise. 

Expanding windows are used to produce
five predictives - one exact and four approximate - for the 60 time points in
the evaluation period, with draws of $\theta$ updated only yearly in all
cases. Draws from $p(\theta,{V|}y\obs)$, used to estimate the \textit{exact} predictive, are produced using a particle Metropolis Hastings (PMH) scheme \citep{AD2010}. The four approximate predictives are produced using the
auxiliary likelihood-based ABC approach of \cite{martin2019auxiliary} { to specify the summary statistics in Algorithm \ref{alg1:rejABC.alg}}. Four alternative models from the generalized autoregressive
conditionally heteroscedastic (GARCH) family of volatility models are chosen
to define the auxiliary likelihood: a GARCH model with normal errors, a GARCH model with Student's t errors, an asymmetric GARCH
with normal errors, and an asymmetric GARCH model with Student's t
errors. We refer the reader to Chapter 9 of \cite{Brooks14} for a
description of these, and related, models; we simply highlight here that such
conditionally deterministic volatility models are suitable for an auxiliary
likelihood-based version of ABC, given the closed-form nature of their
likelihoods. A nearest-neighbour version of
rejection ABC is used, with $N=34992$ and a selection probability of 
$0.86\%$. See \cite{frazier2018} for an explanation of
the rule adopted to determine these values for a given sample size, $n.$

For each predictive, $M=2000$ draws are used to produce an optimal weight, as per (\ref{alp_opt}), the associated optimal portfolio, and its utility. The 60 values of utility are then averaged for a particular predictive method, with five such numbers produced.

\begin{figure}[ptbh]
\centering
\scalebox{0.8}{\input{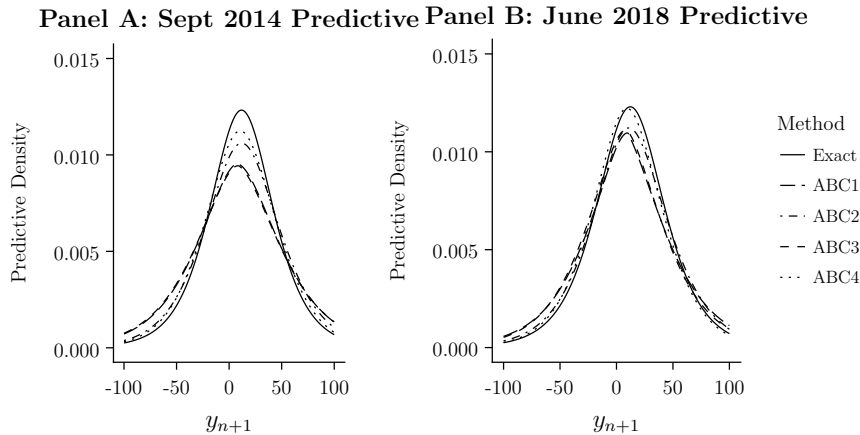}} \caption{One-step-ahead predictive densities for
September 2014 (Panel A) and June 2018 (Panel B): exact (PMH-based) and
approximate (ABC-based).}%
\label{predplot}%
\end{figure}

{Before discussing the resulting (estimated}) {expected utilities, we
present representative exact and approximate predictives in {Panels A and B of
Figure \ref{predplot}}, for two particular months. We label the predictive estimated using PMH as `Exact' and the approximate predictives constructed using each of these four auxiliary models listed above as: `ABC1', `ABC2', `ABC3' and `ABC4', respectively. The plots demonstrate that there is very little to distinguish
between the approximate and exact predictive methods, in particular for June 2018. The similarity between the exact and approximate predictives is further borne
out in the expected utility calculations. For a fixed risk aversion
parameter of $\gamma=4$, the exact approach yields expected utility of
$-4.72$, while the four approximate methods yield expected utilities of
$-4.69,$ $-4.70,$ $-4.69$ and $-4.69$. Moreover, the computation time
required to produce the exact expected utility is ten times larger than
the time required to produce the expected utility via the \textit{slowest}
of the approximate methods, and fifteen times greater than the computation
time for the \textit{fastest} approximate method. A comprehensive set of expected utility estimates were produced, for
different degrees of risk aversion, with the same negligible difference
between exact and approximate results being in evidence.}

{\color{black}We close this section by noting that the close match of the exact and approximate predictives is not simply an artifact of the particular predictive model chosen. In \cite{frazier2019approximate} (cited earlier), comparable numerical results are documented for a range of different models, including for both continuous and discrete data. Moreover, under the required regularity conditions for Bayesian consistency of both the exact and ABC posteriors, the one-step-ahead exact and ABC-based predictives are shown to merge in the sense that, for a large enough sample size, and for models of \textit{any} fixed dimension, the predictive distributions are identical.}

\section{Discussion}
ABC and similar likelihood-free inference methods are emerging as an important part of the analysis toolbox in various application areas where we are able to simulate realistic data, but the models are too complicated for likelihood-based inference. Here we have demonstrated various aspects of using this approach in challenging real-world applications that go beyond the typical benchmark examples used in the literature.

 A particularly interesting combination of inference and generative modeling is the possibility of learning the parameters of a system based on observed data and simulating the performance under various scenarios, treatments or interventions. For example, one of our case studies involved predicting a patient-specific cancer treatment outcome, which could in the future be an important part of treatment planning, especially as the mathematical modeling of disease evolution under treatments is rapidly improving \citep{kozlowska2018}.
A similar application is found in policy planning for epidemics, as the models of infection control are improving, and different `what-if'-scenarios can be explored. Such activities have recently been extensively performed across the world during the Covid-19 pandemic and the iteratively improving understanding about the epidemiology of the virus illustrates the need for proper representations of uncertainties in the model components.

As the modeling proficiency increases, we will be faced with even more difficult inference tasks in higher dimensional domains, with computationally heavier simulators and, in some applications, even requirements for carrying out the inference (and prediction) in real-time. These challenges call for further advances from the statistics and computation community on both the inference algorithms and their efficient implementation via user-friendly software. 

\section*{Acknowledgments}
This work was supported by the European Research Council grant number 742158, European Union’s Horizon 2020 Research and Innovation Programme, Grant Agreement number 847912, Research Council of Norway, Grant Numbers 237718, 311188 and 309273, the Academy of Finland grant number 1316602,  Australian Research Council Discovery Grants DP170100729 and DP200101414 and Australian Research Council Early Career Researcher Award DE200101070.

\bibliographystyle{abbrvnat}
\bibliography{references.bib}

\end{document}